\documentclass{czjphys}         
\begin{document}
\parskip 5pt
\newcommand{\g}{{\bf g}}
\newcommand{\h}{{\bf h}}
\def\nn{\nonumber}
\def\d{\partial}
\def\v{\vert}
\def\l{\langle}
\def\r{\rangle}
\def\beq{\begin{equation}}
\def\eeq{\end{equation}}
\def\bea{\begin{eqnarray}}
\def\eea{\end{eqnarray}}
\def\ba{\begin{array}}
\def\ea{\end{array}}
\title{Bound and scattering states of extended Calogero
model with an additional PT invariant interaction 
}
%
\authori {B. Basu-Mallick, Tanaya Bhattacharyya,  Anjan Kundu} 
\addressi {Theory Group, 
Saha Institute of Nuclear Physics, 
1/AF Bidhan Nagar, Kolkata 700 064, India}
\authorii{Bhabani Prasad Mandal}     \addressii{
Department of Physics, 
Maulana Azad College,
8 Rafi Ahmed Kidwai Road,
 Kolkata 700013, India.}
\authoriii{}    \addressiii{}
\authoriv{}     \addressiv{}
\authorv{}      \addressv{}
\authorvi{}     \addressvi{}
%
\headauthor{B. Basu-Mallick, T. Bhattacharyya, A. Kundu \& B. P. Mandal} 
\headtitle{Bound \& Scattering states of extended Calogero model 
...}             
\lastevenhead{B. Basu-Mallick , T. Bhattacharyya, A. Kundu \& B. P. Mandal
 } 
\refnum{A}
\daterec{XXX}    
\issuenumber{0}  \year{2003}
\setcounter{page}{1}
\maketitle
\begin{abstract}
Here we discuss two many-particle quantum systems, 
which are obtained by adding
some nonhermitian but PT (i.e. combined parity and time reversal)
invariant  interaction to the Calogero model with and without confining
potential. It is shown that the energy eigenvalues
are real for both of these quantum systems. For the case of 
extended Calogero model with confining potential, we obtain
discrete bound states satisfying generalised exclusion 
statistics. On the other hand, the
extended Calogero model without confining
term gives rise to scattering states with continuous spectrum.
The scattering phase shift for this case is determined through the
exchange statistics parameter. We find that, unlike the case of usual 
Calogero model, the 
exclusion and exchange statistics parameter differ from each other in the
presence of PT invariant interaction.
\end{abstract}
\noindent\section{Introduction}
\medskip
Exactly solvable 
 many particle quantum mechanical systems with long-range interactions 
have recently attracted a lot of interest due to their close connection 
with diverse subjects like fractional statistics, random matrix theory, 
level statistics for disordered systems, Yangian algebra etc. 
The $A_{N-1}$ Calogero model (related to $A_{N-1}$ Lie algebra)
is the simplest example of such a dynamical model,  containing $N$
particles on a line and with Hamiltonian given by
\cite {ca}
\beq
H= -  {1\over 2} \sum_{j=1}^N {\d^2 \over \d x_j^2}
+   {\omega^2\over 2}  \sum_{j=1}^N x_j^2 +
   {g \over 2} \sum_{j\neq k} {1 \over (x_j -x_k)^2} \, ,
\label{1}
\eeq
where $g$ is the coupling constant associated with long-range interaction.
One can exactly solve this Calogero model and find out the complete set of
energy eigenvalues as
\beq
E_{n_1,n_2, \cdots , n_N} =
 {N \omega \over 2} \left [ 1 + (N-1)\nu \right ] + \omega \sum_{j=1}^N n_j .
\label {2}
\eeq
Here $n_j$s are non-negative integer valued
quantum numbers with $n_j \leq n_{j+1}$ and $\nu $
is a real positive parameter which is related to $g$ as
\beq
g = \nu^2 - \nu \, .
\label {3}
\eeq
It may be noted that, apart from a constant shift for all energy levels,
the spectrum (\ref {2}) coincides with that of $N$ number of free
bosonic oscillators. Furthermore,  one can easily  remove the above mentioned
 constant shift for all energy levels and express (\ref{2})
 exactly in the form of energy eigenvalues for free oscillators:
$E_{n_1,n_2, \cdots , n_N} = {N \omega \over 2}  + \omega
\sum_{j=1}^N \bar n_j,
$
 where ${\bar n_j} = n_j + \nu ( j-1)  $ are quasi-excitation numbers.
However it is evident that these ${\bar n_j} $s
   are no longer integers and they
satisfy a modified selection rule given by
  ${\bar n_{j+1}} - {\bar n_j}  \geq  \nu $,
 which restricts the difference
between the quasi-excitation numbers to be at least
$ \nu$ apart. As a consequence, the
 Calogero model (\ref{1}) provides a microscopic
realization for generalised exclusion statistics (GES) \cite{ha}
 with $\nu $ representing
the corresponding GES parameter \cite{is,po,bk}.

The Calogero model in
absence of confining potential, i.e. setting $\omega =0 $ in
eqn.(\ref{1}),  is also studied in Ref. \cite{ca}. 
Unlike the earlier case, the spectrum of this model is continuous
and only scattering states occur. Due to such scattering, 
particle momentums in a outgoing $N$-particle plane 
wave get rearranged (reversely ordered)
in terms of momentums in the incoming plane wave.   
The corresponding scattering phase shift is given by 
$ \theta_{sc}= \pi\nu\frac{N(N-1)}{2}$, which 
is simply $\nu \pi$ times the total number of two-body exchanges that 
is needed for rearranging $N$ particles in the reverse order. 
Thus it is natural to identify $\nu$ as the exchange statistics parameter
in this case \cite{po}. It may be noted that this exchange statistics
parameter coincides with the exclusion statistics parameter as defined
earlier in the presence of confining potential.

Recently, theoretical investigations on
different nonhermitian Hamiltonians have received a major
boost because many such systems,
whenever they are invariant under combined parity and time reversal
(PT) symmetry, lead to either real 
  or pairs of complex conjugate energy eigenvalues
\cite{pt1,pt2,pt3,ali,wei}.
Such property of energy eigenvalues in nonhermitian PT invariant
 systems can be related to 
 the pseudo-hermiticity \cite{ali} or anti-unitary symmetry \cite{wei}
of the corresponding Hamiltonians in a general way and 
 to the ODE/IM correspondence for some special cases \cite{ddt}. 
However, as concrete examples of PT symmetric quantum mechanics, 
the Hamiltonians of only one particle
in one space dimension have been usually considered in
the literature so far. 
Therefore it should be interesting to consider nonhermitian
but PT invariant Hamiltonian 
 for N-particle system in one space dimension which remain invariant under
the $PT$ transformation \cite{bm}   
\beq 
i\rightarrow -i, \ \ x_j \rightarrow - x_j, \ \  p_j \rightarrow p_j \, ,
\label {4}
\eeq
where $j \in [1,2, \cdots , N],$  and
$x_j$ ($p_j \equiv -i \frac {\partial }{\partial  x_j}$) denotes
the coordinate (momentum) operator of the $j$-th particle.
In particular, one may construct an extension of Calogero model with or
without confining term by
adding to it some nonhermitian but PT invariant interaction,
and enquire whether such extended model would lead to
real spectrum.

The aim of the present article is to shed some light 
on the above mentioned issue for some special cases, where 
the PT invariant extension of the Calogero model can be solved exactly.  
In Sec.2 of this article we consider such a PT invariant
extension of $A_{N-1}$ Calogero model \cite{bk,bm} and show that,
within a certain range of the related parameters, this
extended Calogero model yields real energy eigenvalues obeying GES.
In Sec.3 we consider PT invariant
extension of Calogero Model without confining potential
 and calculate the corresponding scattering phase shift \cite{sca}.
 Section 4 is the concluding section. 

\noindent \section {Bound states of extended Calogero model with 
confining interaction}

Let us consider a nonhermitian but PT invariant extension of the  
Hamiltonian  (\ref{1}) as 
\beq
 {\cal H} = H +  \delta \sum_{j\neq k} {1 \over x_j - x_k}{\d \over \d x_j}
 \, ,
\label {5}
\eeq
where $\delta $ is a real parameter. It may be noted that, 
  Calogero models and their distinguishable variants
 have been solved recently
by mapping them to a system of free oscillators \cite{gp}.
With the aim of solving the extended Calogero model (\ref {5})
by similar method, 
we assume that (justification for this assumption will be given later)
the corresponding ground state wave function 
 is given by
\beq
\psi_{gr}= e^{- {\omega\over 2}  \sum_{j=1}^N x_j^2 }
\prod_{j<k} (x_j-x_k)^\nu  ,
\label {6}
\eeq
where $\nu $ is a real positive number which is
 related to the coupling constants
$g$ and $\delta $ as
\beq
g = \nu^2 - \nu ( 1+ 2 \delta ) \, .
\label {7}
\eeq
Now if we use the expression
 (\ref {6}) for a similarity transformation to
 the Hamiltonian (\ref {5}), it reduces to an `effective Hamiltonian'
 of the form
\beq
{\cal H}'=\psi_{gr}^{-1} {\cal H} \psi_{gr}  = S^- + \omega S^3 + E_{gr} \, ,
\label {8}
\eeq
where the Lassalle operator ($S^-$) and Euler operator ($S^3$) are given by
\beq
 S^- = -  {1\over 2} \sum_{j=1}^N {\d^2 \over \d x_j^2} -
  (\nu- \delta)\sum_{j \neq k}
  {1 \over x_j - x_k}{\d \over \d x_j}, ~~~
  S^3 =  \sum_{j=1}^N  x_j { \d \over \d x_j } ,
\label {9}
\eeq
and
\beq
E_{gr}= {N \omega \over 2} \left [ 1 + (N-1)(\nu -\delta ) \right ] .
\label {10}
\eeq
It is easy to see that
 the Lassalle operator and Euler operator, as defined
 in eqn.(\ref {9}), satisfy the simple
commutation relation:
$ [S^3,S^-]= -2 S^-$. Using therefore the well known Baker-Hausdorff
transformation we can remove the $S^-$ part of the effective Hamiltonian
${\cal H}' $ and through some
additional similarity transformations reduce it finally to the
free oscillator model \cite{bk}
\beq
H_{free} = {\cal S}^{-1} \left({\cal H}'- E_{gr}\right) {\cal S}
  = -  {1\over 2} \sum_{j=1}^N {\d^2 \over \d x_j^2} +
 {\omega^2 \over 2}  \sum_{j=1}^N x_j^2 - {\omega N \over 2},
\label {11}
\eeq
where ${\cal S} =e^{ {1\over 2 \omega } S^- }
e^{ {1\over 4 \omega } \nabla^2 }
e^{ {\omega \over 2} \sum_{j=1}^N x_j^2 }  $. As a consequence 
of these similarity transformations, 
nonsingular eigenfunctions of
  the extended Calogero model (\ref {5}) can be
obtained from the
eigenfunctions of free oscillators as
\beq
 \psi_{n_1, n_2, \cdots , n_N} = \psi_{gr} \ {\cal S} \Lambda_+
\left\{ \prod_{j=1}^N
e^{ - {\omega \over 2} x_j^2 } H_{n_j}(x_j) \right \} \, ,
\label {12}
\eeq
where $H_{n_j}(x_j)$ denotes the Hermite polynomials of
order $n_j$ and $ \Lambda_+ $ 
projects the distinguishable many-particle wave functions to the
bosonic part of the Hilbert space by completely symmetrising
all coordinates. Evidently, the eigenfunctions
(\ref {12}) will be mutually independent if the excitation
numbers $n_j$s obey the bosonic selection rule: $n_{j+1} \geq n_j $.
The eigenvalues of the Hamiltonian (\ref{5})
 corresponding to the states  (\ref{12}) will
  naturally be given by 
\beq
E_{n_1,n_2, \cdots , n_N} = E_{gr}+\omega \sum_{j=1}^N n_j=
 {N \omega \over 2} \left [ 1 + (N-1)(\nu -\delta ) \right ] +
 \omega \sum_{j=1}^N n_j .
\label {13}
\eeq
It is worth noting that, for the purpose of obtaining real
 eigenvalues (\ref {13}) as well as nonsingular eigenfunctions
  (\ref {12})  at the limit $x_i
\rightarrow x_j $, 
$\nu $ should be taken as a real positive parameter. Due to eqn.(\ref {7}),
  this condition restricts the ranges of coupling
constants $g$ and $\delta $ as 
(i) $\delta > - {1\over 2} , ~~ 0 > g >  - (\delta +{1\over 2})^2  $,
and (ii) $g>0$ with arbitrary value of $\delta $.
Thus the energy eigenvalues (\ref {13})
 of the PT invariant Hamiltonian (\ref {5}) would be real within the 
above mentioned ranges of the coupling constants. 
 Furthermore, it is evident that for all $n_j=0,$
the energy $E_{n_1,n_2, \cdots , n_N}$
  attains its minimum value
$E_{gr}$. At the same time, as can be easily seen 
 from eqn.(\ref {12}),
 the corresponding  eigenfunction  reduces to
$\psi_{gr}$ (\ref {6}).
 
To explore the GES in the case of PT invariant model (\ref {5}),
we observe that  eqn.(\ref{13}) can be rewritten \cite{bk}
exactly in the form of energy spectrum for $N$ free oscillators as
\beq
E_{n_1,n_2\cdots n_N} = \frac{N \omega }{2} + \omega
\sum_{j=1}^N \bar{n}_j \, ,
\label{14}
\eeq
where $
\bar{n}_j = n_j +  (\nu - \delta ) (j-1) $.  These
 quasi-excitation numbers ($\bar{n}_j$)
evidently  satisfy a modified selection rule:
$ \bar{n}_{j+1}- \bar{n}_j \geq  \nu - \delta \,  .$  Since the minimum
 difference between two consecutive
$\bar{n}_j $s is given by
\beq
 \tilde{\nu } = \nu - \delta_ ,
\label {16}
\eeq
the spectrum of extended Calogero model (\ref {5})
satisfies GES with parameter $ \tilde{\nu}$. 


Since both $A_{N-1}$ Calogero model (\ref {1}) and its nonhermitian
extension  (\ref {5}) can be solved by mapping them  
to a system of free harmonic oscillators,
it is natural to enquire whether these
models are directly related through some similarity transformation.
Investigating along this line, we find that
\beq
\Gamma ^{-1}{\cal H} \Gamma =  H^\prime =
  \frac{ 1}{2}\sum p_j^2 + \frac{ 1}{2}\omega ^2
\sum x_j^2  + g^\prime
\sum_{j\neq k}^N \frac{1}{(x_j-x_k)^2} \, ,
\label{17}
\eeq
where  $ \Gamma = \prod_{j< k} (x_j-x_k)^\delta  $,
and $H^\prime $ denotes the Hamiltonian of $A_{N-1}$
Calogero model with  `renormalised' coupling constant
 given by $g^\prime = g + \delta (1+\delta ) $. However, 
due to the above mentioned similarity transformation, 
 singular eigenfunctions of the usual Calogero model (\ref {1})
can generate nonsingular 
eigenfunctions of the extended Calogero model (\ref {5}) 
in some region of the parameter space \cite{bm,tia}.
As a result, the spectrum of extended Calogero model differs
qualitatively from the spectrum of the original Calogero model and 
leads to a negative value of the GES parameter (\ref {16})
in such region of parameter space.

\noindent \section {Scattering states of extended Calogero model 
without confining interaction}

Here our aim is to study the 
PT invariant extension of $A_{N-1}$ Calogero model in the absence of 
confining interaction. Putting $\omega =0$ in ${\cal H}$ (\ref {5}),
 we explicitly obtain the Hamiltonian of such extended Calogero model as
\beq
 {\cal H }_0
= -\frac{1}{2}\sum_{j=1}^N \frac{\partial^2}{\partial x_j^2} 
+ \frac{g}{2}
\sum_{j\neq k}\frac{1}{(x_j - x_k)^2} + \delta \sum_{j \neq k} \frac{1}
{(x_j - x_k)}\frac{\partial}{\partial x_j}\, .
\label{18}
\eeq
Following \cite{ca}, we try to solve the eigenvalue problem 
corresponding to the above Hamiltonian within a sector of configuration 
space corresponding to a definite ordering of particles like 
$x_1 \geq x_2 \geq \cdots \geq x_N$.
We find that the solutions of the
eigenvalue equation ${\cal H}_0\psi = p^2\psi$, 
where $p$ is real and positive, are given by \cite{sca}
\beq
\psi = \prod_{j<k}{(x_j-x_k)}^{\nu}P_{k,q}(x)r^{-b}J_b(pr) \,  .
\label{19}
\eeq 
Here the `radial' coordinate $r$ is
defined as: $r^2 = \frac{1}{N}\sum_{i<j}{(x_i-x_j)}^2$,  
  $b$ is given by 
$ b=k+\frac{(N-3)}{2}+ \frac{1}{2}N(N-1)\tilde {\nu}$, and the
 parameters $\nu , \, {\tilde \nu}$ 
are defined exactly in the same way as in section 2. 
Moreover, $J_b(pr)$ denotes the Bessel function and $P_{k,q}(x)$s are
translationally invariant, symmetric, $k$-th order homogeneous polynomials
satisfying the differential equations
\beq
\sum_{j=1}^N \frac{\partial^2 P_{k,q}(x)}{\partial x_j^2} 
+ \tilde{\nu}\sum_{j \neq
k}\frac{1}{(x_j-x_k)}\left(\frac{\partial}{\partial x_j} - \frac{\partial}
{\partial x_k}\right) P_{k,q}(x) = 0\, .
\label{20}
\eeq
Note that 
the index $q$ in $P_{k,q}(x)$ can take any integral value ranging 
from $1$ to $g(N,k)$, where $g(N,k)$ is the
 number of independent polynomials 
which satisfy eqn.(\ref{20}) for a given $N$ and $k$ \cite{ca}. 

It is evident that, within the same range of coupling constants 
for which ${\cal H}$ (\ref {5}) yields discrete bound states 
with real energy eigenvalues, ${\cal H}_0$ (\ref {18}) yields continuum 
scattering states with real energy eigenvalues. 
Due to (\ref{19}), the most general eigenfunction for ${\cal H}_0$ with
eigenvalue $p^2$ can be written as
\beq
\psi = \prod_{j<k}{(x_j-x_k)}^{\nu}\sum_{k=0}^{\infty}\sum_{q=1}^{g(N,k)}
C_{kq} r^{-b} J_b(pr) P_{k,q}(x)\, ,
\label{21}
\eeq
where $C_{kq}$s are some arbitrary constants.
To discuss scattering, we need only the asymptotic behaviour of
the wavefunction (\ref{21}) when all particles are far apart from each
other. Hence, using the asymptotic properties of Bessel function at $r
\rightarrow \infty $ limit, one can
write $\psi$ (\ref {21}) as 
\beq
\psi \sim \psi_{+} + \psi_{-}
\label{22}
\eeq
where
$$
\psi_{\pm} = {(2\pi pr)}^{-\frac{1}{2}}\prod_{j<k}{(x_j-x_k)}^
{\nu}r^{-A}\sum
_{k=0}^{\infty}\sum_{q=1}^{g(N,k)}C_{kq}r^{-k}P_{k,q}(x)
e^{\pm i(b + \frac{1}{2})\frac{\pi}{2}\mp ipr}. 
$$
By choosing the coefficients $C_{kq}$ in a
proper way, the incoming wavefunction ($\psi_{+}$) can be 
expressed in the form of a plane wave like 
\beq
\psi_{+}= C\exp[\,i\sum_{j=1}^N p_j x_j\,]  \, ,
\label{23}
\eeq
where
$p_j \leq p_{j+1}$, \ \ 
$p^2 =\sum_{j=1}^N p_j^2 $
and
$\sum_{j=1}^N p_j = 0. $
Then, by following the approach of \cite{ca}, it can be shown that the
outgoing wavefunction ($\psi_{-}$) 
 takes the form \cite{sca}
\beq
\psi_{-} = Ce^{-i\pi\nu\frac{N(N-1)}{2}}\exp[\,i\sum_{j=1}^N x_{j}
\,p_{N+1-j}\,] \, .
\label{24}
\eeq
Comparing (\ref {23}) with (\ref {24}), we find that
 the momentums of incoming plane wave gets rearranged (reversely ordered)
in the scattering 
process and the corresponding phase shift is given by $\pi\nu
\frac{(N(N-1)}{2}$. Thus $\nu$ can be identified with the exchange 
statistics parameter associated with this phase shift. 

\noindent\section {Conclusion}

Here we have constructed  two exactly solvable
many-particle quantum systems by adding
some nonhermitian but PT 
invariant  interaction to the $A_{N-1}$
Calogero model with and without confining
potential. It is shown that the energy eigenvalues
are real for both of these quantum systems in some region of the
parameter space. The exclusion statistics parameter for the case of 
extended Calogero model with confining potential is determined through 
the allowed energy levels of discrete bound states.
 On the other hand, the exchange statistics parameter
for the case of extended Calogero model without confining
term is determined through the 
 scattering phase shift of plane waves. 
 Surprisingly we find that, 
in contrary to the case of original Calogero model, 
 the exclusion and exchange statistics parameters
 derived in the above mentioned way differ from each other in the
presence of PT invariant interaction. As a future study, it might be 
interesting to find out the inner product for which the eigenstates 
of extended Calogero models would satisfy the orthonormality property 
and completeness relation. 

\noindent {\bf Acknowledgments }
 
One of the authors (BBM) would like to thank
Prof. M. Znojil for kind invitation and hospitality
during the `1st International Workshop 
on Pseudo-Hermitian Hamiltonians in Quantum Physics', 
Prague, June 16-17, 2003.

\end{document}